\documentstyle[psfig,conf_iap]{article}

\def\etal{{\it et al.}}

\def \lleq {\lower0.9ex\hbox{ $\buildrel < \over \sim$} ~}
\def \ggeq {\lower0.9ex\hbox{ $\buildrel > \over \sim$} ~}
\def \rb{\bar{r}}
\def \sb{\bar{s}}
\def\l{\Lambda}

\def\spose#1{\hbox to 0pt{#1\hss}}
\def\simle{\mathrel{\spose{\lower 3pt\hbox{$\mathchar"218$}}
\raise 2.0pt\hbox{$\mathchar"13C$}}}
\def\simge{\mathrel{\spose{\lower 3pt\hbox{$\mathchar"218$}}
     \raise 2.0pt\hbox{$\mathchar"13E$}}}
\def\apj{{\it Astroph.~J.~}}

\def\aj{{\it Astron.~J.~}}

\def\beq{\begin{equation}}
\def\eeq{\end{equation}}
\def\ber{\begin{eqnarray}}
\def\eer{\end{eqnarray}}
\newcommand{\sq}{\lower.25ex\hbox{\large$\Box$}}

\begin{document}

\heading{Exploring dark energy using the Statefinder}
\par\medskip\noindent

\author{
Varun Sahni~$^1$
}

\address{Inter-University Centre for Astronomy \& Astrophysics,
Pun\'e 411 007, India}

\begin{abstract}
Observations of high redshift supernovae indicate that the universe is 
accelerating. The hypothesis of `Dark energy' 
(cosmological constant,
scalar field tracker potentials, braneworld models,
etc.) has been advanced to explain this phenomenon.
Sensitive tests of dark energy which can differentiate between
rival models are clearly the need of the hour.
The statefinder pair $\lbrace r,s\rbrace$ is a geometrical diagnostic which
can play this role. $r$ \& $s$ depend upon the third time derivative of
the scale factor $\stackrel{...}{a}$ and provide the next logical step 
in the hierarchy of the cosmological parameter set after $H$ and $q$.
The statefinder pair $\lbrace r,s\rbrace$ can be determined to high
accuracy from a SNAP type experiment and allows us to successfully
differentiate between dark energy models
having constant as well as time-varying
equations of state.
\end{abstract}

\bigskip

Type Ia supernovae, treated as standard candles,
lead to a compelling scenario in which the universe is accelerating,
driven by a form of `dark energy' which could have large negative pressure
\cite{perlmutter99,riess98}.
Many candidates for dark energy have been advanced 
\cite{ss00,sahni02} and several are
in good agreement with current observational constraints. 
Rapid advances in the observational situation and prospects for the
deployment of a space-telescope dedicated to supernova research (SNAP)
add momentum to our quest for determining the nature of dark energy.
Clearly the need of the hour is to combine high-z Sn searches with 
diagnostic tools which are sensitive to the properties of dark energy
and could thereby help in distinguishing between rival models.
The Statefinder, recently introduced in \cite{sssa02}, promises to do
just that. The Statefinder pair $\lbrace r,s\rbrace$ is a diagnostic of
dark energy which is constructed from second and third derivatives 
of the scale factor. As a result it contains information about both
the equation of state $w$ as well as its time derivative ${\dot w}$.

Geometrical parameters have traditionally played a key role in cosmology.
The first of these is the Hubble 
parameter $H_0 = ({\dot a}/a)_0$ and the second is the deceleration
parameter $q_0 = - H_0^{-2}({\ddot a}/a)_0$. In order to be able to 
differentiate between different forms of dark energy we add to this
hierarchy a third geometrical parameter
\beq
r = \frac{\stackrel{...}{a}}{a H^3}.
\label{eq:state1a}
\eeq
For models of dark energy with equation of state $w$ and density parameter
$\Omega_X$, the first Statefinder $r$ has the form
\beq
r = 1 + \frac{9 w}{2}\Omega_X (1+w)
- \frac{3}{2}\Omega_X \frac{\dot w}{H}.
\label{eq:state1b}
\eeq
Combining $r$ and $q$ leads to the second Statefinder 
\beq
s = \frac{r - 1}{3(q - 1/2)} \equiv 1 + w - \frac{1}{3}\frac{\dot w}{w H}.
\label{eq:state2}
\eeq
An interesting property of $s$ is that it does not explicitly depend upon the 
dark energy density $\Omega_X$.

\noindent
\begin{figure}[tbh!]
\centerline{
\psfig{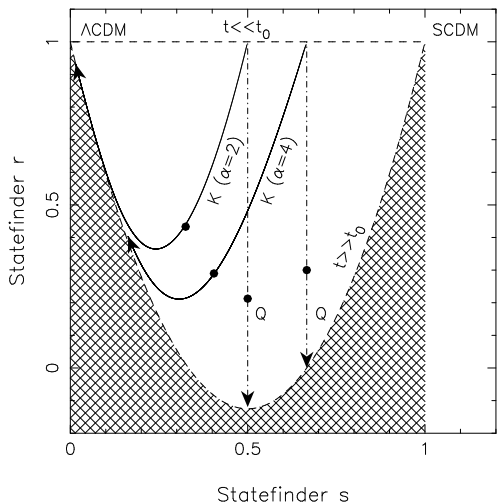} }
\caption{\small The Statefinder pair ($r,s$) for different
models of dark energy. Quiessence (Q) models have a constant equation of state
($w = $constant $\neq -1$). In them
the value of $s$ remains fixed at $s = 1+w$ while the value of $r$
asymptotically declines to
$r(t \gg t_0) \simeq 1 + \frac{9w}{2}(1+w)$.
Two models of Quiessence corresponding to $w_Q = -0.25, -0.5$ are shown.
Kinessence (K) is a scalar field rolling down
the potential $V(\phi) \propto \phi^{-\alpha}$ with $\alpha = 2,4$.
K models commence their evolution on a tracker trajectory (at $t \ll t_0$)
and asymptotically approach $\Lambda$CDM at late times.
$\Lambda$CDM ($r=1,s=0$) and SCDM without radiation ($r=1,s=1$) are fixed
points of the system.
The hatched region is disallowed in Quiessence models and in the
Kinessence model which we consider.
The filled circles show the {\em current values} of the Statefinder pair
($r_0,s_0$) for the Q and K models ($\Omega_{0m} = 0.3$).
Reproduced from \cite{sssa02}.
}
\label{fig:state}
\end{figure}

The Statefinder pair $\lbrace r, s \rbrace$
has several useful properties. 
For instance $\lbrace r, s \rbrace$ = $\lbrace 1,0 \rbrace$ in a universe containing 
a cosmological
constant and non-relativistic matter ($\l$CDM) while
$\lbrace r, s \rbrace$ = $\lbrace 1,1 \rbrace$ 
for the standard cold dark matter model containing 
no radiation. 

As illustrated in figure \ref{fig:state}, the fixed point 
$\lbrace r, s \rbrace$ = $\lbrace 1,0 \rbrace$ is a {\em late time attractor} for
a large class of dark energy models including those in which acceleration
is caused by the slow roll of a scalar field down a `tracker potential'.
However the current value 
$\lbrace r_0, s_0 \rbrace$ in tracker models can differ significantly from
$\lbrace 1,0 \rbrace$ and it is this feature which
allows us to successfully differentiate
between different tracker models of dark energy. 
Figure \ref{fig:state} also shows `Quiessence' models
for which the equation of state remains constant as the universe expands.
These models follow vertical trajectories in the plane defined by 
$\lbrace r(t), s(t) \rbrace$ and do not approach the $\l$CDM
fixed point at late times. Figure \ref{fig:state} clearly shows that
models with evolving as well as unevolving equations of state can be
easily differentiated by the Statefinder statistic.

An important property of the statefinder diagnostic is that it is geometrical
in nature since it depends upon the expansion factor and its derivatives.
This distinguishes it from physical descriptors of dark energy including
$\Omega_X$, (or $T_0^0$ and $T_\alpha^\alpha$).
Geometrical and physical parameters are related to one another through
the field equations of cosmology; a summary is given in table 1.

\begin{table}[tbh!]
\begin{center}
\caption{Relationship between geometrical and physical parameters
characterizing the observable Universe}
\bigskip
\begin{tabular}{ll}
\hline
Geometrical parameters & Related physical parameters\\
\hline
$H = {\dot a}/a$ & $\Omega_{\rm total}, \Omega_{\rm curvature}$\\
$q = -{\ddot a}/aH^2$ & $\Omega_i, w_i$\\
$r = {\stackrel{...}{a}}/{a H^3}$ & $\Omega_i, w_i, {\dot w_i}$\\
$s = (r - 1)/3(q - 1/2)$ & $w_i, {\dot w_i}$\\
$r_c = \int dt/a$ & $\Omega_i, w_i$\\
\hline
\end{tabular}
\label{table:lambda}
\end{center}
\end{table}
\bigskip


\noindent
\begin{figure}[tbh!]
\centerline{
\psfig{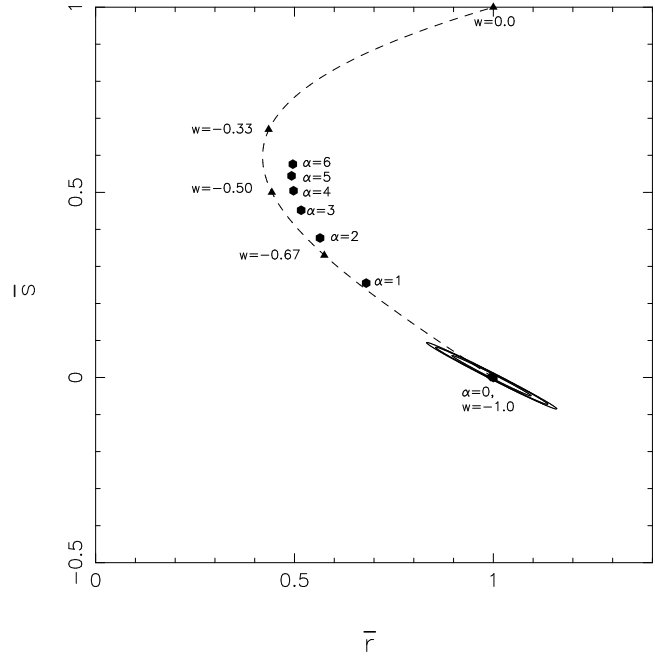} }
\caption{\small
$1\sigma$, $2\sigma$, $3\sigma$ confidence levels of $\rb$ and
$\sb$ computed from $1000$ SNAP-type experiments.
The fiducial model is $\Lambda$CDM
with $\Omega_{0m} = 0.3, \Omega_{0\Lambda} = 0.7$.
Filled circles
show values of $\rb$ and $\sb$ for the `tracker'
potential $V(\phi) \propto \phi^{-\alpha}$ with
$\alpha=1,2,3,4,5,6$ (bottom to top).
Quiessence models ($w$ = constant $\neq -1$) lie on the dashed line,
filled triangles on this line show
$w = -2/3, -1/2, -1/3, 0$
(bottom to top). We note that all inverse power-law models (and most
Quiessence models) lie
well outside of the three sigma contour centered
around the $\Lambda$CDM model and so does the braneworld model
\cite{DDG} for which $\bar {r} = 0.7, \bar {s} = 0.27$. 
Reproduced from \cite{sssa02}.
}
\label{fig:snap}
\end{figure}

Figure \ref{fig:snap} shows confidence levels for the `mean 
statefinders'
\beq
\bar{r} = \frac{1}{z_{\rm max}}\int_0^{z_{\rm max}} r(z)\,dz\,\,,~~
\bar{s} = \frac{1}{z_{\rm max}}\int_0^{z_{\rm max}} s(z)\,dz.
\eeq
computed from a SNAP-type experiment which probes a $\l$CDM fiducial model
with $\Omega_{0m} = 0.3, \Omega_{0\Lambda} = 0.7$
(see \cite{sssa02} for details).
The discriminatory power of the Statefinder diagnostic is demonstrated by the
fact that inverse power law models of dark energy
lie well outside of the three sigma
contour centered around $\l$CDM.

Finally we would like to emphasize that the Statefinder diagnostic can be applied 
with advantage to an
extremely general category of models including several for which the notion
of equation of state is not directly applicable. These include 
models in which acceleration is caused by a change in
the theory describing gravitation (hence space-time expansion) and not
due to the presence of a fluid with negative pressure; examples include
braneworld models \cite{DDG,shtanov,alam} and scalar-tensor theories  
\cite{scalar-tensor}. 
Interesting results using Statefinders have recently been obtained 
for the Chaplygin gas \cite{kamen}.

\acknowledgements{The work reported here was done in collaboration with
Alexei Starobinsky, Tarun Deep Saini and Ujjaini Alam to whom I would like
to express my gratitude.}

\vfill
\end{document}